\documentclass[aps,prl,twocolumn,showpacs,floatfix]{revtex4}
\usepackage{graphicx}
\usepackage{dcolumn}
\usepackage{bm}
\usepackage{epsf}
\usepackage{subfigure}
\usepackage{epstopdf}
\usepackage{amsmath}
\usepackage{amssymb}

\newcommand{\e}{{\rm e}}

\begin{document}

\title{Thermoelectric efficiency at maximum power in a quantum dot}

\author{Massimiliano Esposito}
\altaffiliation[]{Also at Center for Nonlinear Phenomena and Complex Systems,
Universit\'e Libre de Bruxelles, Code Postal 231, Campus Plaine, B-1050 Brussels, Belgium.\\}
\author{Katja Lindenberg}
\affiliation{Department of Chemistry and Biochemistry and Institute for Nonlinear Science, 
University of California, San Diego, La Jolla, CA 92093-0340, USA}
\author{Christian Van den Broeck}
\affiliation{Hasselt University, B-3590 Diepenbeek, Belgium}

\date{\today}

\begin{abstract}
We identify the operational conditions for maximum power of a nanothermoelectric engine consisting of a single quantum level 
embedded between two leads at different temperatures and chemical potentials.
The corresponding thermodynamic efficiency agrees with the Curzon-Ahlborn expression up to quadratic terms in the gradients,
supporting the thesis of universality beyond linear response. 
\end{abstract}

\maketitle

The purpose of this letter is to present a detailed thermodynamic analysis of
electron transport  through a single quantum dot connecting two leads at different temperatures and 
chemical potentials.  
Of particular interest to us is the efficiency of the thermal motor function, in which electrons are
pumped upward in chemical
potential under the impetus of a downward temperature gradient. 
The study of this model addresses several issues of timely interest: nanotechnology, the study of
thermodynamic properties of small devices that are prone to fluctuations, the question of universality for
thermodynamic properties away from equilibrium, the role of quantum features in this respect, and the
promise of thermoelectricity generated by nano-devices. We briefly comment on each of these topics.

Spectacular technical and experimental progress in nano- and biotechnology
 have greatly increased our ability to observe, manipulate, control, and even manufacture
systems on a very small scale~\cite{Bustamante,Muller}. In parallel, new theoretical tools and concepts have been
developed that make it possible to exhibit the deeper relationship between fluctuations,
entropy production and work, and the role of stochasticity in small scale non-linear
nonequilibrium phenomena. In particular, stochastic thermodynamics formulates the thermodynamics of
small entities subject to thermal fluctuations~\cite{Schnakenberg,VandenBroeck1,se,Seifert05,Esposito07}.
These developments are closely related to the celebrated fluctuation~\cite{fluctuationtheorem} 
and work theorems~\cite{worktheorem}. 

The concept of Carnot efficiency is a central cornerstone of thermodynamics. According to this principle,
the efficiency, defined as the ratio of work output over heat input for a machine operating between two
thermal baths at temperatures $T_l$ and $T_r$ ($T_r>T_l$) is at most equal to
\begin{equation}
\eta_c=1-\frac{T_l}{T_r}.
\end{equation}
The equality is only reached for reversible operation. This is a universal result
which remains valid for small scale fluctuating systems such as the well-documented case of Brownian
motors, see~\cite{VandenBroeck6} and references therein. However, reversible processes require
infinitely slow operation, implying that such engines produce zero power.  
One of the important questions, when operating away from equilibrium, is the efficiency at maximum power. 
In a groundbreaking paper, Curzon and Ahlborn~\cite{curzon} calculated this efficiency for the Carnot 
engine in the so-called endo-reversible approximation (taking into account the dissipation only in 
the heat transfer process). They found a strikingly simple formula, 
namely 
\begin{eqnarray}\label{CA}
\eta_{CA}=1-\sqrt{1-\eta_c}\approx \eta_c/2+\eta_c^2/8+6 \eta_c^3/96+\ldots \;. 
\end{eqnarray}
Recently, it has been shown that the  Curzon-Ahlborn efficiency is an exact consequence of 
\emph{linear} irreversible thermodynamics when operating under conditions of strong coupling 
between the heat flux and the work~\cite{VandenBroeck5}.  The value of $1/2$ for the linear 
coefficient in Eq. (\ref{CA}) is therefore universal for such systems.

The efficiency at maximum power  was also addressed in the context of stochastic 
thermodynamics in~\cite{Seifert07b}, where it was shown that the efficiency at 
maximum power for a Brownian particle undergoing a Carnot cycle through the modulation of a harmonic 
potential is given by $\eta_S=2\eta_c/(4-\eta_c)\approx \eta_c/2+\eta_c^2/8+3 \eta_c^3/96+\ldots$. 
By an entirely different calculation, dealing with the Feynman ratchet and pawl model (which operates 
under steady rather than cyclic conditions), the efficiency at maximum power was found to be~\cite{Tu} 
$\eta_T=\eta_c^2/[\eta_c-(1-\eta_c)\ln(1-\eta_c)]\approx \eta_c/2+\eta_c^2/8+7 \eta_c^3/96+\ldots$. 
All three of the above results agree, as they should~\cite{VandenBroeck5}, to linear order in
$\eta_c$.  More surprisingly, the coefficient of the quadratic term is also identical.
This raises the question as to whether universality also applies to the coefficient of the quadratic term.

The field of thermoelectricity went through a revival in the early 1990s due to the discovery of new 
thermoelectric materials with significantly higher thermodynamic yields~\cite{Snyder08}. 
Of particular interest are the developments in the context of nanostructured materials~\cite{Majumbar04}. 
For example, thermoelectric experiments have been reported on silicon nanowires~\cite{Heath07}, 
individual carbon nanotubes~\cite{Kim03} and molecular junctions~\cite{Majumdar07}. 
Furthermore, it has been reported that Carnot efficiency can be reached
for electron transport between two leads at different temperatures and chemical potentials, by
connecting them through a channel sharply tuned at the energy for which the electron density
is the same in both leads~\cite{Linke02,Linke05}. 
A double-barrier resonant tunneling structure has been proposed 
as a possible technological implementation~\cite{Linke06}. 

\begin{figure}[t]
\centering
\rotatebox{0}{\scalebox{0.7}{\includegraphics{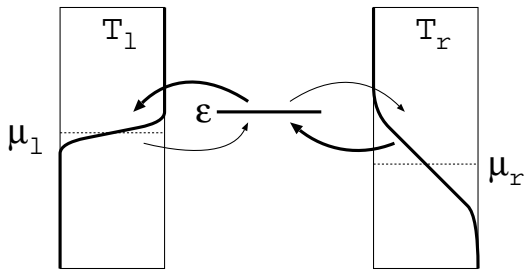}}}
\caption{Sketch of the nanothermoelectric engine consisting of a 
single quantum level embedded between two leads at different temperatures 
and chemical potentials. We choose by convention $T_l<T_r$. 
Maximum power is observed in the regime  $\varepsilon>\mu_l>\mu_r$.}
\label{plot0}
\end{figure}

The thermoelectric device whose properties we explore below is arguably the simplest prototype of such systems. 
It consists of a quantum dot with a single resonant energy level in contact with two thermal reservoirs at different 
temperatures, see Fig.~\ref{plot0}. The dot can contain one single electron with a sharply defined energy $\varepsilon$.  
The exchange of electrons between the leads through the dot will be described by a stochastic master 
equation~\cite{Bonet,Nazarov,EspositoHarbola1}, and the corresponding thermodynamic properties can be 
obtained from stochastic thermodynamics~\cite{Schnakenberg,VandenBroeck1,se,Seifert05,Esposito07}. 
In anticipation of the forthcoming analysis, we note that this model displays perfect coupling
between energy and matter flow: because of the sharply defined dot energy, every single electron 
carries exactly the same amount of energy. Hence,  Carnot efficiency will be  reached when operating close
to equilibrium~\cite{VandenBroeck6,Linke02,Linke05}, while  Curzon-Ahlborn efficiency will be
obtained at maximum power in the regime of linear response~\cite{VandenBroeck5}. 
Going beyond these results, we will identify the operational conditions for working at maximum power. 
In particular, the efficiency at maximum power will be found to be
$\eta \approx \eta_c/2+\eta_c^2/8+\ldots$, with the coefficient of $\eta_c^2$ again equal to 
$1/8$. This provides further support for the thesis of universality for this value, 
especially since the regime of maximum power is found to lie entirely in the quantum regime.

We now turn to the mathematical analysis of the thermoelectric engine represented 
in  Fig.~\ref{plot0}. A single level quantum dot, with orbital energy $\varepsilon$, 
exchanges electrons with a cold left lead, temperature $T_l$ and chemical potential $\mu_l$, and 
with a hot right lead, temperature $T_r$ and chemical potential $\mu_r$. 
The quantum dot is either empty (state $1$) or filled (state $2$). 
The crucial variables of the problem are the scaled energy barriers (with $k_B=1$)
\begin{equation}
x_\nu = \frac{\varepsilon-\mu_\nu}{T_\nu}, \quad \nu=l,r.
\label{scaledenergy}
\end{equation}
The exchange of electrons with the leads is described by 
the following quantum master equation~\cite{EspositoHarbola1,Nazarov,Bonet}:
\begin{eqnarray}
\left(
\begin{array}{c}
\dot{p}_{1}(t) \\ 
\dot{p}_{2}(t) 
\end{array}
\right) =
\left(
\begin{array}{cc}
- W_{21} & W_{12} \\
W_{21} & - W_{12}
\end{array}
\right)
\left(
\begin{array}{c}
p_{1}(t) \\ 
p_{2}(t) 
\end{array}
\right) \;. \label{MaterEq2by2}
\end{eqnarray} 
The rates are given by
\begin{eqnarray}
W_{12} &=& \sum_{\nu=l,r} W_{12}^{(\nu)} = \sum_{\nu=l,r} a_{\nu} (1-f_{\nu}) \\
W_{21} &=& \sum_{\nu=l,r} W_{21}^{(\nu)} = \sum_{\nu=l,r} a_{\nu} f_{\nu} ,\label{ModelRates}
\end{eqnarray}
where $f_{\nu}=[\exp(x_\nu)+1]^{-1}$ is the Fermi distribution.  
The fact that $a_{\nu}$ is independent of the dot energies is known as the wide band approximation. 

We are interested in the properties of the device at the steady state. 
The steady state distributions for the dot occupation follow from  
$W_{21} p_{1}^{ss}= W_{12} p_{2}^{ss}$ with $p_1^{ss}+p_2^{ss}=1$. 
The resulting probability current from the lead $\nu$ to the dot is then
\begin{eqnarray} 
{\cal I}_{\nu} \equiv W_{21}^{(\nu)} p_{1}^{ss}-W_{12}^{(\nu)} p_{2}^{ss} .
\label{DefHeatExtrnu}
\end{eqnarray}
Using ${\cal I}_{r}=-{\cal I}_{l}$ and $W_{12}+ W_{21}= a_r+a_l$, 
we can rewrite the result for the flux from the right lead as 
\begin{eqnarray} \label{Ir}
{\cal I}_{r}= \alpha (f_r-f_l) ,
\end{eqnarray}
where $\alpha=a_r a_l/(a_r+a_l)$. Eq.~(\ref{Ir}) is essentially the Landauer formula for a single channel.

The steady state heat per unit time $\dot{{\cal Q}}_{r}$ extracted from the lead $r$, and the steady state 
work per unit time (power) $\dot{{\cal W}}$ performed by the device upon bringing 
electrons from right to left lead, are respectively given by:
\begin{eqnarray} 
\label{DefHeatExtr}
\dot{{\cal Q}}_r &=&(\varepsilon-\mu_{r}) {\cal I}_{r}
= \alpha T_{r} x_r (f_r-f_l) \\
\dot{{\cal W}}&=&(\mu_l-\mu_r) {\cal I}_{r} 
= \alpha T_r \big(x_r - (1-\eta_c) x_l \big) (f_r-f_l)\;.\nonumber\\
&&
\label{DefWorkExtr}
\end{eqnarray}  
The corresponding thermodynamic efficiency reads 
\begin{eqnarray} 
\eta \equiv \frac{{\cal W}}{{\cal Q}_{r}} 
= \frac{\dot{{\cal W}}}{\dot{{\cal Q}}_{r}} 
= \frac{\mu_l-\mu_r}{\varepsilon-\mu_r} 
= 1-(1-\eta_c) \frac{x_l}{x_r}\label{DefEffi}.
\end{eqnarray}

The entropy production associated with the master equation (\ref{MaterEq2by2}) 
is given by~\cite{Schnakenberg,VandenBroeck1,Seifert05,Esposito07}
\begin{eqnarray} 
\sigma = \sum_{i,j,\nu} W_{ij}^{(\nu)} p_{j}^{ss} 
\ln \frac{W_{ij}^{(\nu)} p_{j}^{ss}}{W_{ji}^{(\nu)} p_{i}^{ss}} \geq 0,
\end{eqnarray}
where $i,j=1,2$.
Noting that $\ln [W_{12}^{(\nu)}/W_{21}^{(\nu)}]=x_{\nu}$, one finds, 
in agreement with standard irreversible thermodynamics~\cite{GrootMazur}, 
the following expression for the entropy production:
\begin{eqnarray} \label{ep}
\sigma = F_m J_m + F_e J_e = \alpha (x_l-x_r) (f_r-f_l) \geq 0,
\end{eqnarray}
with thermodynamic forces for matter and energy flow, $F_m$ and $F_{e}$, given by
\begin{equation} 
F_m \equiv -(\frac{\mu_r}{T_r}-\frac{\mu_l}{T_l}), \quad
F_e \equiv \frac{1}{T_r}-\frac{1}{T_l}.
\end{equation}
We stress that the corresponding matter and heat flow, given by
\begin{equation} 
J_m \equiv- {\cal I}_{r}, \quad
J_e \equiv- \varepsilon {\cal I}_{r} , 
\end{equation}
are proportional to each other. In other words, matter and heat flow are perfectly coupled and
the condition for attaining both Carnot and Curzon-Ahlborn efficiency, namely, that the determinant
of the corresponding Onsager matrix be zero, is fulfilled~\cite{VandenBroeck6,VandenBroeck5}.\\

We first discuss the case of equilibrium. Due to the perfect coupling, it is enough to stop one
current, matter or energy, and the other one will automatically vanish. Under this condition, detailed
balance is valid, ${\cal I}_{\nu}=0$.
It is clear from  Eq.~(\ref{Ir}) that the matter flux (and hence also the energy flux) vanishes
if and only if $f_{l}=f_{r}$ or, equivalently, $x_{l}=x_{r}$. The 
efficiency then becomes equal to Carnot efficiency, cf. Eq.~(\ref{DefEffi}), and the entropy
production vanishes, $\sigma=0$ [cf. Eq.~(\ref{ep})]. Note that $x_{l}=x_{r}$ does not require that
the thermodynamic forces $F_m$ and $F_{e}$ vanish separately, i.e.,
at this singular balancing point equilibrium does not require temperature and chemical potential to be
identical in both reservoirs~\cite{Linke02,VandenBroeck6,VandenBroeck5,Linke05}.\\

We next turn to the operational condition for maximum power. For given temperatures $T_l$ and $T_r$,
we search for the values of the scaled electron energy barriers $x_l$ and $x_r$ that maximize $\dot{{\cal W}}$. From
$\partial_{x_l} \dot{{\cal W}}=\partial_{x_r} \dot{{\cal W}}=0$,
we find the following two equations determining these values:
\begin{eqnarray} 
\label{2equ1}
(f_l-f_r)+\left[x_r-(1-\eta_c) x_l\right]f_r^2 \e^{x_r}=0 \\
(f_l-f_r)+\big(\frac{x_r}{1-\eta_c}- x_l\big) f_l^2 \e^{x_l}=0 .
\label{2equ2}
\end{eqnarray}
A first observation is that  these equations depend only on the ratio of the two temperatures. 
Second, while the equations involve transcendental relations, one obtains the following explicit
result by subtracting the first equation from the second,
\begin{eqnarray} 
\label{isolatexl}
x_l = 2 \ln \left[ \frac{\cosh \{ x_r/2 \}}{\sqrt{1-\eta_c}} 
+ \sqrt{ \frac{\cosh^2 \{ x_r/2 \}}{1-\eta_c}  -1} \right].
\end{eqnarray}
Substitution of this result in (\ref{2equ1}) gives
\begin{eqnarray}
&&x_r-\sqrt{2} \cosh (x_r/2) \sqrt{2 \eta_c-1 + \cosh (x_r)} 
+2(\eta_c-1) \nonumber\\&&\hspace{0.5cm} \times \ln \left[ \frac{\cosh (x_r/2)
+\sqrt{2 \eta_c-1+\cosh (x_r)}/\sqrt{2}}{\sqrt{1-\eta_c}} \right] \nonumber \\
&&\hspace{0.5cm}+ \cosh (x_r)=0. \label{BigEq}
\end{eqnarray}
Since an analytic solution of this equation is not possible, we first turn to perturbative solutions
for $\eta_c$ close to the limiting values $0$ (reservoirs of equal temperatures) and $1$ (cold
reservoir at zero temperature).  For the case $\eta_c \rightarrow 0$, we substitute
$x_r=a_0+a_1 \eta_c + a_2 \eta_c^2 + {\cal O}(\eta_c^3)$
in Eq.~(\ref{BigEq}) and expand the resulting equation in $\eta_c$.
The coefficients $a_1$, $a_2$, etc., are found recursively by solving order by order in  $\eta_c$. 
At order zero, we find an identity.
At first order, we find the transcendental equation
$a_0 = 2 \coth (a_0/2)$.
The numerical solution is $a_0=2.39936$.  
At second order and  third order in $\eta_c$, we find
$a_1=-a_0/4$ and
$a_2= \sinh (a_0)/\left\{6\left[1-\cosh (a_0)\right]\right\}$.
Substitution of these results
and (\ref{isolatexl}) in (\ref{DefEffi}) leads to the following expansion of the
efficiency at maximum power in the regime of small $\eta_c$:
\begin{eqnarray} 
\label{AnalyExpEta}
\eta=\frac{\eta_c}{2}+\frac{\eta_c^2}{8}
+\frac{\left[7+{\rm csch}^2 (a_0/2)\right]}{96} \eta_c^3 + {\cal O}(\eta_c^4) \;.
\end{eqnarray}
The expansion features the expected coefficient $1/2$ for the linear term, 
but also supports the thesis that the coefficient of $\eta_c^2$ has a 
universal value, namely, $1/8$.

We next turn to the analysis of the case $\eta_c \rightarrow 1$ ($T_l \to 0$). 
As we will see by self-consistency, the value of $x_r$ converges 
to a finite limiting value. 
With this a posteriori insight, we can easily identify the leading 
behavior of $x_l$ from Eq. (\ref{isolatexl}), namely,
$x_l\sim -\ln(1-\eta_c)$.
Substitution of this result in (\ref{2equ1}) leads to the conclusion that  
$x_r \rightarrow b$, where $b$ is a solution of the transcendental equation 
\begin{eqnarray} \label{teb}
e^{-b}+1=b, \label{xr1}
\end{eqnarray}
with numerical solution $b=1.27846$. The corresponding efficiency converges to $1$, albeit rather
slowly, cf. Eq.~(\ref{DefEffi}). 
\begin{figure}[h]
\centering
\rotatebox{0}{\scalebox{0.44}{\includegraphics{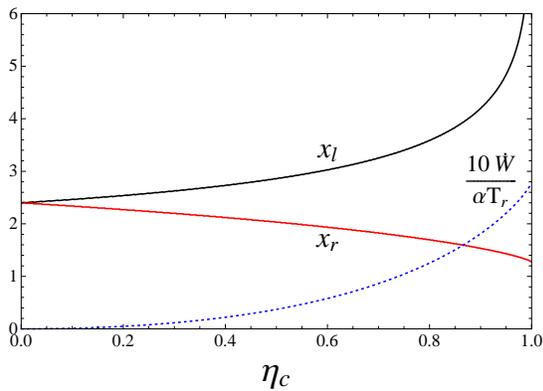}}}
\caption{(Color online) Scaled electron energy barriers $x_l$ and $x_r$  at maximum power, cf. Eqs.~(\ref{isolatexl}) 
and (\ref{BigEq}), as  a function of the Carnot efficiency $\eta_c=1-T_l/T_r$. 
The dotted line represents the corresponding (scaled) power $10 \times \dot{{\cal W}}/(\alpha T_r)$, cf. Eq.~(\ref{DefWorkExtr}).}
\label{plot1}
\end{figure}
\begin{figure}[h]
\centering
\rotatebox{0}{\scalebox{0.44}{\includegraphics{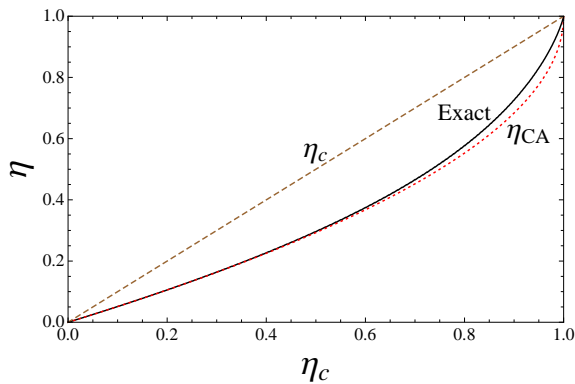}}}
\caption{(Color online) Efficiency at maximum power in a single level quantum dot as a function of the 
Carnot efficiency $\eta_c=1-T_l/T_r$ (full line), as compared to Curzon-Ahlborn efficiency (dotted line)
and Carnot efficiency (dashed line).}
\label{plot2}
\end{figure}
To complete the picture, we show in Fig.~\ref{plot1} the numerical solutions for $x_l$ and $x_r$ as
functions of $\eta_c$ as well as the corresponding rescaled maximum power $\dot{{\cal W}}/(\alpha T_r)$. 
Note that  $x_r$ is always of order unity, so that the regime of maximum power
cannot be well described by either a high or a low temperature expansion. 
It can also be seen that the maximum power is a monotonically increasing function of
$\eta_c$. The corresponding efficiency is reproduced in Fig.~\ref{plot2}. It is very close to
the Curzon-Ahlborn efficiency, with relative deviations largest for large $\eta_c$.  

In conclusion, nanosystems with perfectly coupled fluxes, such as the quantum dot described here,
are of great interest.  They can operate as steady state Carnot engines. They probably possess universal
features up to quadratic terms in nonlinear response when working at maximum power. One can speculate
that they offer, from a technological point of view, advantages over their macroscopic counterparts.
The above analysis can be repeated for the quantum dot operating as a refrigerator, corresponding to
the regime $x_r\geq x_l$. Such an analysis reveals that maximum cooling power (maximum $\dot{{\cal Q}}_{r}$
extracted from the cold lead) is attained for $x_r \rightarrow \infty$ and $x_l \rightarrow b$, where
$b$ is again the solution to the transcendental equation (\ref{teb}). The corresponding cooling
power is thus $b T_l$ per transported particle, to be compared with the cooling power
of $T_l$ for a classical engine.

\section*{Acknowledgments}

M. E. is supported by the FNRS Belgium (charg\'e de recherches) and 
by the Luxembourgish government (Bourse de formation recherches). 
This research is supported in part by the
National Science Foundation under grant PHY-0354937.

\vskip 10pt


\end{document}